\documentclass[runningheads]{svmult}

\usepackage{makeidx}   
\usepackage{graphicx}  
\usepackage{subeqnar}  
\usepackage{multicol}  
\usepackage{physprbb}  
\makeindex             

\begin{document}
\title*{The Mechanism of Core-Collapse Supernova \protect\newline
Explosions: A Status Report\footnote{Support for this work was provided by
the Scientific Discovery through Advanced Computing (SciDAC) program
of the DOE, grant number DE-FC02-01ER41184, and
by NASA through Hubble Fellowship
grant \#HST-HF-01157.01-A awarded by the Space Telescope Science
Institute, which is operated by the Association of Universities for
Research in Astronomy, Inc., for NASA, under contract NAS 5-26555.}} 
\toctitle{The Mechanism of Core-Collapse Supernova \protect\newline
Explosions: A Status Report} 
%
%
\titlerunning{Neutrino-Driven Supernova Mechanism}
%
\author{Adam Burrows\inst{1}
\and Todd A. Thompson\inst{2,3}}
\authorrunning{Adam Burrows and Todd A. Thompson}
\institute{The University of Arizona, Tucson, AZ 85721
\and The University of California,
     Berkeley, CA 94720
\and Hubble Fellow} 

\maketitle              

\section{Introduction}

Most massive stars (8 M$_{\odot}$ $\rightarrow$ 80 M$_{\odot}$) 
must transition at the ends of their lives into neutron
stars or stellar-mass black holes.  That they do so when their low-entropy cores 
reach the Chandrasekhar mass, gravitationally collapse, and launch a supernova explosion 
has been demonstrated both by direct observations (cf.,  the neutrinos 
from SN1987A) and by a host of compelling theoretical arguments.  However,
numerical simulations of the process of core collapse, bounce near nuclear densities,
shock wave generation, and shock propagation have failed to recreate in detail
the observed gravitational masses of known neutron stars, the expected nucleosynthesis pattern,
and empirical supernova energies.  All semi-realistic, one-dimensional (spherically-symmetric)
simulations conducted to date fizzle into quasi-static accreting proto-black holes.
What is more, they seem to do so convincingly, despite concerted attempts over
the years to include all the known neutrino and nuclear physics or general relativity
\cite{bruenn.1985,myra,bruennandmezz97,bruenn2001,mezz2001,lieben2001,lieben2002,rampp2000}. 
It is now fairly clear that the devil is not ``in the details'' and that
1D models don't explode.  Something large and major, not something at 
the ``10-20\% level,'' would have to be missing to alter this conclusion.  

Two-dimensional simulations conducted in the nineties\cite{herant,bhf.1995,janka.muller96,fryer}
and more recent three-dimensional SPH simulations\cite{fryerwarren}
do explode (when the corresponding simulations in one 
dimension do not), but these multi-D numerical experiments do not
produce supernova explosions that satisfy all of the above observational constraints.    
Furthermore, they all employ some realization of a simple flux-limited, 
energy-integrated diffusion algorithm, in lieu of full neutrino transport (which
is quite difficult in 2- or 3-D). It is thought by many that the apparent 
marginality of success demands that the transport be handled with a multi-group,
multi-angle technique (and that the hydrodynamics be handled in full general relativity)
before the multi-D explosion simulations are to be considered 
valid, or even indicative of the mechanism.  Perhaps.  However, this puts 
too much reliance on hardware and software, and not enough on physical understanding.

\bigskip
So, what is wrong?  And how to fix it?  In order to see where we are going
we must first see where we have been.

\section{One-Dimensional Simulations: The Spherically-Symmetric Paradigm}
\label{1D}

The primary elements of the neutrino-driven mechanism were mapped out 
in the work of Colgate and White\cite{cw} and Arnett\cite{arnett}.  In the former,
a cold (low-entropy), already deleptonized core was collapsed and after bounce at nuclear densities
a neutrino luminosity which heated the outer mantle was turned on for 14 milliseconds.
The neutrino transport was artificial, but it drove an explosion with $5\times 10^{52}$ ergs.
In the latter, a higher-entropy core was collapsed, which bounced due to thermal nucleon pressure at a central
density near a tenth of nuclear, and neutrino transport was handled more realistically with 
an energy-integrated diffusion approach.  Neutrino diffusion was found to transport energy
from the hot interior to the mantle on close to the dynamical timescale of the newly-generated
shock wave.  In Arnett's calculations, this kept the shock wave from stalling, an explosion
ensued, and the total duration of the neutrino luminosity was less than 100 milliseconds, 
but longer than in Colgate and White.  In both cases, neutrinos played the 
central role in launching the explosion and a neutron star remained, but 
the initial progenitor models, neutrino transport, neutrino physics,
and equations of state were primitive by today's standards.  The timescale for the cooling phase
(detected for SN1987A) was off by a factor of 100-1000, the spectra were much harder than 
current simulations indicate, and ``$\nu_{\mu}$'' neutrinos were not included. 
Nevertheless, the core idea of the neutrino-driven mechanism, that neutrinos generated
in the inner core and deposited in the outer mantle drive the explosion, was implicit in both papers.   

The essential ingredient of the neutrino-driven mechanism is the transfer of
gravitational binding energy from the core to the mantle of the newly-formed protoneutron star
through the mediation of neutrinos liberated at the high temperatures and densities 
generated in its interior.  An {\it efficient} coupling is required, that, if not achieved, 
will lead to a fizzle.  With up-to-date nuclear equations of state, neutrino physics, progenitor models,
and transport algorithms, the coupling efficiency in one-dimensional simulations 
has been determined to be inadequate both to forestall the stagnation
of the bounce shock and to reignite it once it has stalled.  Figure \ref{masst} depicts radius trajectories
of mass zones versus time for one of our 1D multi-group, Feautrier/tangent-ray calculations (\S\ref{Burr}).
The progress of the shock and its stagnation during the first 250 milliseconds after
bounce are manifest.   
\begin{figure}[b]
\vspace*{3in}
\caption[]{Radius trajectories of selected mass zones in our 300-zone, 20 energy-group
calculation of stellar collapse, bounce, and shock propagation for the Woosley and Weaver
11 M$_{\odot}$ progenitor\cite{woosley.weaver}.  In this simulation, as in all other 1D simulations, the shock does not revive.}
\label{masst}
\end{figure}
Wilson\cite{wilson} and Bethe and Wilson\cite{bethe}
suggested that shock stagnation is only temporary, leading to the ``delayed" scenario, 
but without the neutron-finger convective boost (an inherently multi-D 
effect which others have failed to reproduce) in the driving neutrino 
luminosity Wilson himself did not find explosions.
In sum, embellishments with general relativity, the employment of multi-group, multi-angle
Boltzmann solvers, and refinements in neutrino-matter interactions have not changed the
conclusion that in spherical symmetry the coupling efficiency of the emergent neutrinos to the protoneutron
star mantle is too small to lead to explosion, even after a 
pause\cite{bruenn.1985,myra,bruennandmezz97,bruenn2001,mezz2001,lieben2001,lieben2002,rampp2000}.  
Unfortunately, this conclusion seems to be independent of progenitor model.

\subsection{Supernova Energetics Made Simple}

It is important to note that one is not obliged to unbind the inner core ($\sim$10 kilometers) as well; the 
explosion is a phenomenon of the outer mantle at ten times the radius (50-200 kilometers).  
One consequence of this goes to the heart of a general confusion concerning 
supernova physics.  Though the binding energy of a cold neutron star is $\sim$$3 \times 10^{53}$ ergs
and the supernova explosion energy is near $10^{51}$ ergs, a comparison of these two numbers and
the large ratio that results are not very relevant.  More germane are the binding energy
of the mantle (interior to the shock or, perhaps, exterior to the neutrinospheres)
and the neutrino energy radiated during 
the delayed phase.  These are both at most a few$\times$10$^{52}$ ergs, not
$\sim$$3 \times 10^{53}$ ergs, and the relevant ratio that illuminates
the neutrino-driven supernova phenomenon is $\sim$10$^{51}$ 
ergs divided by a few$\times$10$^{52}$ ergs. This is $\sim$5-10\%, not 
the oft-quoted 1\%, a number which tends to overemphasize the 
sensitivity of the neutrino mechanism to neutrino and numerical details.       

Furthermore, there is general confusion concerning what determines the supernova explosion energy.
While a detailed understanding of the supernova mechanism is required to
answer this question, one can still proffer a few observations.  First is the simple
discussion above.  Five to ten percent of the neutrino energy coursing through
the semi-transparent region is required, not one percent.  Importantly, the optical
depth to neutrino absorption in the gain region is of order $\sim$0.1.  The product
of the sum of the $\nu_{e}$ and $\bar{\nu}_e$ neutrino energy emissions in the first
100's of milliseconds and this optical depth gives a number near 10$^{51}$ ergs.
Furthermore, the binding energy of the progenitor mantle exterior to the iron core
is of order a few$\times 10^{50}$ to a few$\times 10^{51}$ ergs and it is very approximately
this binding energy, not that of a cold neutron star, that is relevant in setting the 
scale of the core-collapse supernova explosion energy.  Given the power-law nature
of the progenitor envelope structure, it is clear that this binding energy is related
to the binding energy of the pre-collapse iron core (note that they both have a boundary
given by the same $GM/R$), which at collapse is that of
the Chandrasekhar core.  The binding energy of the Chandrasekhar core is easily
shown to be zero, modulo the rest mass of the electron times the number of baryons
in a $\sim$1.4 M$_{\odot}$ Chandrasekhar mass.  (The Chandrasekhar mass/instability is tied to the 
onset of relativity for the electrons, itself contingent upon the electron rest mass).  
The result is $\sim$10$^{51}$ ergs.

The core-collapse explosion energy is near the explosion energy for a Type Ia supernovae because in a thermonuclear explosion
the total energy yield is approximately the 0.5 MeV/baryon derived from carbon/oxygen burning to
iron times the number of baryons burned in the explosion.  
The latter is $\ge$half the number of baryons in a Chandrasekhar
mass.  The result is $\sim$10$^{51}$ ergs.  This is the same number as for core-collapse supernovae because
1) in both cases we are dealing with the Chandrasekhar mass (corrected for electron captures,
entropy, general relativity, and Coulomb effects) and 2) the electron mass and the per-baryon
thermonuclear yield are each about 0.5 MeV.

While more detailed calculations are clearly necessary to do this correctly, the essential
elements of supernova energetics are not terribly esoteric (if neutrino-driven), at least to within a factor of 5,
and should not be viewed as such.

\subsection{Some Recent Results using our Feautrier/Tangent-Ray/ALI Method}
\label{Burr}

One is firm in the general conclusion that 1D models with good physics and numerics
do not explode because, coming from many different angles, various groups have now verified this. 
We here briefly describe our contribution to this activity.

It is thought that the angular distribution of neutrinos in the gain region
needs to be calculated with precision, since neutrino energy deposition is
proportional to neutrino energy density and, for given luminosities/fluxes,
this is sensitive to $\nu_e$ and $\bar{\nu}_e$ angular distributions that can not be derived using
flux-limited diffusion.  Furthermore, the stiff dependence of the absorption
cross sections on neutrino energy requires multi-group approaches.  Hence,
multi-group, multi-angle Boltzmann transport algorithms are to be preferred.  

We have constructed such a code, using the Feautrier variables advocated 
in standard stellar atmospheres work\cite{mihalas80} and the tangent-ray method
to establish a dense angular grid\cite{burrows.2000}.  Our transport solver calculates in the comoving frame,
uses accelerated $\Lambda$ iteration (ALI)\cite{can73a,can73b,scharmer81,olson}
to speed convergence of the solution, is 
implicit in time, second-order accurate in space, and iterates between the Boltzmann/transfer
equation and the zeroth- and first-moment equations (in a method akin to the variable Eddington
factor approach) until a converged global solution to the full transport equation is achieved.  No ad hoc flux limiters or
artificial closures are necessary.  This iteration scheme is fast (convergence to a part in 10$^6$ in 2 to 10 steps) 
and automatically conserves energy in the transport sector.  
The Feautrier scheme can transition to the diffusion
limit seamlessly and accurately and the tangent-ray method automatically adapts with
the hydrodynamic grid as it moves.  In constructing the tangent rays, we cast 
them from every outer zone to every inner zone.  Hence, if there are 200 radial zones,
the outer zone has 199 angular groups.  Because of the spherical nature of the core-collapse
problem and the need to accurately reproduce the angular distribution of the radiation
field that transitions from the opaque (inner) to the transparent (outer) regions,
such fine angular resolution is useful, though computationally 
demanding, as radiation becomes more and more forward-peaked.
Figure \ref{angle} depicts a snapshot of the angular distribution of the $\nu_e$ neutrinos
at various energies from 1 to 320 MeV, at a radius of 42 kilometers, 40 milliseconds after
bounce.  The positions of the angular bins are shown as dots in the lower hemisphere.
The progressively more forward-peaked distribution at lower energies (less coupled)
is clearly well-resolved.  The code calculates the Feautrier variables (and, hence, the
specific intensity) at every radial zone, for every energy group, for each neutrino
species (we follow the standard ``3"), at each timestep.
\begin{figure}[b]
\vspace*{3in}
\caption[]{A snapshot of the angular distribution of the $\nu_e$ specific intensity at a radius
of 42 kilometers, 40 milliseconds after bounce, for energy groups from 1 to 320 MeV.
The 11 M$_{\odot}$ progenitor of Woosley and Weaver\cite{woosley.weaver} was used.}
\label{angle}
\end{figure}
We employ 20-40 energy groups either logarithmically or linearly spaced
from 1 MeV to either 100 MeV (for $\bar{\nu}_{e}$ and ``$\nu_\mu$'') or 320 MeV (for $\nu_e$).

The hydrodynamics is explicit in time and Lagrangean, uses a predictor/corrector method, employs artificial
viscosity to handle shocks, and is Newtonian.  We had originally used a PPM hydro scheme, but could
not easily incorporate neutrino radiation pressure into the Riemann solver. We have constructed a dense equation 
of state table in the variables T, $\rho$, and Y$_e$ that contains the 14 variables 
needed by the radiation hydrodynamics.   We have constructed additional tables of the $\nu_i$-electron
scattering kernels and the $e^+/e^-$ annihilation kernels\cite{bruenn.1985,thomp}.  Neutrino-electron
redistribtion/scattering is handled explicitly, which is a great time and memory saver.  The radiation/matter
couplings (both the energy and electron fraction updates) are handled implicitly and in operator-split 
fashion, using the ALI to facilitate convergence.  Figure \ref{plot_ye} provides snapshots we obtain of
a representative evolution of the Y$_e$ profile before and after trapping and bounce.
\begin{figure}[b]
\vspace*{3in}
\caption[]{A collection of snapshots of the Y$_{e}$ profile versus interior mass for the
Newtonian evolution of the Woosley and Weaver\cite{woosley.weaver} 11 M$_{\odot}$ progenitor. A total
of $\sim$250 milliseconds is depicted and the simulation was carried to $\sim$50 milliseconds after bounce.}
\label{plot_ye}
\end{figure}
 
The minuses of our approach are that it is Newtonian, that the hydro is explicit,
and that, being Lagrangean and not adaptive, the envelope must be pre-zoned densely in mass to maintain reasonable
resolution of the shock at late times.  Rampp and Janka\cite{rampp2000,rampp2002} use an Eulerian grid and 
PPM hydro, remap between the comoving frame radiation solution and their static
hydro grid, employ a variable Eddington factor/tangent-ray/Feautrier scheme as well, and 
operator split the neutrino-matter couplings.  They cast tangent rays to a subset of
interior Eulerian zones and thereby maintain a static, fixed set of angles. In addition, they have 
a simple method for approximately incorporating general relativity.  There are lots of minor
differences between our numerical implementations, but despite 
them our simulation results are quite similar\cite{thomp}.

Liebend\"orfer et al.\cite{lieben2001,lieben2002} have assembled a very different radiation/hydrodynamic code complex
that has both strengths and weaknesses.  Among their many strengths is that their code is fully
implicit, adaptive (a great advantage), and general relativistic.  Minor weaknesses are that 
they operator-split all the terms in the transport equations separately 
and use perforce (due to their implicit redistribution method)
a small number of angles (6) in the S$_n$ method they utilize to handle the angular distribution
of the radiation field.  S$_n$ can not handle forward-peaked radiation fields, but 
for the core collapse problem, the gain region interior to a stalled shock (where the angular
distribution is most problematic) is sufficiently compact that their S$_n$ technique is more
than adequate.  Where it fails, in the outer regions, the neutrino/matter coupling is not so germane
to the problem of the supernova mechanism.  There are many other differences of implementation
and approach, but, again, despite these differences their results for velocities,
trapped electron fractions, entropies, densities, and the effect of neutrino-electron redistribution are both qualitatively 
and quantitatively similar to those of both ourselves\cite{thomp} 
and Rampp and Janka\cite{rampp2000,rampp2002}.

Figure \ref{lum} portrays the evolution of the luminosity of the electron neutrinos
with time up to 200 milliseconds after bounce for three progenitor models.  The breakout
peak, and the precursor peak just a few milliseconds earlier, are not only similar to one another,
but they are similar to the corresponding results of Rampp and Janka\cite{rampp2000}.  Though mostly similar
to the results of Liebend\"orfer et al. off the peak, their peak luminosities can be 20-30\% higher than ours.
Due to the 20-M$_{\odot}$ model's thicker envelope, the subsequent accretion luminosity 
for this progenitor is $\sim$60\% higher than that for the  
two other progenitors 200 milliseconds after bounce.
\begin{figure}[b]
\vspace*{3in}
\caption[]{The electron-neutrino luminosity versus time around bounce for three different progenitors.
The primary focus of this plot is the prodigious shock breakout burst.}
\label{lum}
\end{figure}
In none of our 1D simulations, carried out to as long as $\sim$1 
second after bounce and with the best neutrino and nuclear 
physics available, do the models explode.  Moreover, we have incorporated, in approximate fashion,
the possible effects of the neutrino oscillations, using the numbers derived from  
the solar and atmospheric neutrino experiments, and have seen no effect.  This is not unexpected, since 
the ``matter effect" severely suppresses oscillation between flavor neutrinos.

We are forced to conclude, on the basis of both our simulations 
and those of others, that the program to determine whether 
core-collapse supernovae explode in spherical symmetry (1D) 
when all the best physics and numerics are implemented has 
failed by a comfortable (?) margin. This leads us of necessity to multi-D effects.

\section{Multi-Dimensional Simulations}
\label{2D}

The success of the 2D BHF\cite{bhf.1995} and Fryer et al.\cite{fryer} simulations can be traced to the increase in the efficiency
of the neutrino-matter coupling in the gain region interior to the shock due to the longer dwell time
of convecting/absorbing parcels of matter.  In 1D, accreted matter falls straight through
the gain region.  Consequently, in 1D the increase in entropy due to neutrino energy deposition is modest.
In 2- or 3-D, matter resides a bit longer in the gain region due to convection.  As a result, the steady-state entropy
of an average parcel of matter is higher.  This translates into a larger gain region, that is on average less bound
and more unstable to favorable changes in the global conditions of accretion and luminosity.  In the language of
Burrows and Goshy\cite{goshy}, the global instability condition of the protoneutron star is eased
and changes in the control parameters (such as $\dot{M}$ and L$_{\nu_e}$) can more easily
result in explosion.  Whether they do when better neutrino transfer and general relativity are
implemented is unclear, but recently Janka and Rampp (JR, this volume) have seen a very weak explosion
in a calculation using 2D hydro in a 27$^{\circ}$ wedge and their 1D Feautrier transport along rays.
They obtain no explosion in 1D for the same inputs.
In this 2D calculation, an average hydro solution in the wedge is used as input to a spherical calculation of the Eddington
factors, and these are employed for all the individual Feautrier solutions along rays.  This is better than
the flux-limited solution along rays of BHF, incorporates better neutrino transport,  
has an approximate prescription for relativity, but seems to result in a weaker explosion.  However, 
the JR algorithm is not yet true 2D transport and the calculations need to be done 
over a full 180$^{\circ}$ angular region.  Nevertheless, it is clear from the JR calculation
and those of BHF\cite{bhf.1995}, Herant et al.\cite{herant}, and 
Fryer et al.\cite{fryer} that convection makes the protoneutron
star mantle more unstable.  What is not clear is whether this is enough.

Fryer and Warren\cite{fryerwarren} have recently demonstrated that their 3D SPH simulations explode quantitatively and
qualitatively in much the same way as their 2D simulations, despite a general concern that 2D and 3D convection
have different clump size spectra and cascade character.  This seems not to matter much; it is the
overall global behavior of the convecting region that matters and, according to them, the largest
size and velocity scales.  Small scales are a detail.  The former are set by the 
size of the unstable region and are similar in both
2D and 3D.  However, these calculations need to be verified with multi-group, multi-angle techniques,
first in 2D (which has yet to be done), then in 3D (a major computational challenge for the future).

\section{Coda}
\label{conclusion}

Multi-dimensional effects seem to be required to ensure or enable core-collapse
supernova explosions.  However, even though what the essential elements for explosion are
is unclear, we believe that the solution, when found, will not be marginal. 
A number of groups are now embarked upon separate development
paths to credible 2D (later 3D) radiation/hydro schemes (Arizona/Israel, ORNL, MPA).  We, in collaboration
with Eli Livne and Itamar Lichtenstadt, are developing a 2D, moving grid, 
discontinuous finite element (DFE), multi-group, and multi-angle (S$_n$) neutrino radiation/hydrodynamics code.  
With it, we hope soon to simulate multi-D collapse and to test out the numerous avenues now before us.    
Stay tuned.

%

\end{document}